\documentclass[universe,article,accept,pdftex,moreauthors]{Definitions/mdpi}

\firstpage{1} 
\pubvolume{1}
\issuenum{1}
\articlenumber{0}
\pubyear{2024}
\copyrightyear{2024}
\externaleditor{{Academic Editor: } 
}
\datereceived{} 
\daterevised{} 
\dateaccepted{} 
\datepublished{ } 
\hreflink{https://doi.org/} 

\usepackage{graphicx,epsfig,rotate}
\usepackage[labelformat=simple]{subcaption}

\DeclareCaptionLabelFormat{subcaptionlabel}{\normalfont(\textbf{#2}\normalfont)}
\def\e10{\eta_{10}}

\def\Ombh2{\Omega_{\rm b} h^2}

\def\ga{\mathrel{\raise.3ex\hbox{$>$\kern-.75em\lower1ex\hbox{$\sim$}}}}
\def\la{\mathrel{\raise.3ex\hbox{$<$\kern-.75em\lower1ex\hbox{$\sim$}}}}

\def\beq{\begin{equation}}
\def\eeq{\end{equation}}
\def\beqar{\begin{eqnarray}}
\def\eeqar{\end{eqnarray}}

\newcommand{\trh}{T_\text{RH}}
\newcommand{\arh}{a_\text{RH}}

\DeclareMathOperator{\csch}{csch}

\Title{Inflaton decay in No-Scale Supergravity and Starobinsky-like models \small{\rightline{UMN--TH--4317/24, FTPI--MINN--24/08}}}

\TitleCitation{Inflaton decay in No-Scale Supergravity and Starobinsky-like models}

\Author{Yohei Ema$^1$, Marcos A. G. Garcia$^2$, Wenqi Ke$^1$, Keith A. Olive$^1$,  and Sarunas Verner$^3$}

\AuthorCitation{Ema, Y.; Garcia, M. A. G.; Ke, W.; Olive, K.A.; Verner, S.}

\address{%
$^{1}$ \quad William I. Fine Theoretical Physics Institute,
School of Physics and Astronomy,
University of Minnesota, Minneapolis, MN 55455, USA\\
$^{2}$ \quad Departamento de F\'isica Te\'orica, Instituto de F\'isica, Universidad Nacional Aut\'onoma de M\'exico, Ciudad de M\'exico C.P. 04510, Mexico \\
$^{3}$ \quad Institute for Fundamental Theory, Physics Department, University of Florida, Gainesville, FL 32611, USA}

\corres{Correspondence: olive@umn.edu}

\abstract{
We consider the decay of the inflaton in Starobinsky-like models arising from either an $R+R^2$ theory of gravity or $N=1$ no-scale supergravity models. If Standard Model matter is simply introduced to the $R + R^2$ theory, the inflaton (which appears when the theory is conformally transformed to the Einstein frame) couples to matter predominantly in Standard Model Higgs kinetic terms. This will typically lead to a reheating temperature of $\sim 3 \times 10^9$~GeV. However, if the Standard Model Higgs is conformally coupled to curvature, the decay rate may be suppressed and vanishes for a conformal coupling $\xi = 1/6$. Nevertheless, inflaton decays through the conformal anomaly leading to a reheating temperature of order $10^8$~GeV. The Starobinsky potential may also arise in no-scale supergravity. In this case, the inflaton decays if there is a direct coupling of the inflaton to matter in the superpotential or to gauge fields through the gauge kinetic function. We also discuss the relation between the theories and demonstrate the correspondence between the no-scale models and the conformally coupled $R+R^2$ theory (with $\xi = 1/6$).
}

\begin{document}
\section{Introduction}

A key aspect of the difference between the `old' model of inflation, based on a first-order phase transition \cite{Guth:1980zm}, and the `new' inflationary Universe \cite{new} is the ability to recover from the era of exponential expansion, or a graceful exit from inflation \cite{reviews}.  In old inflation, bubble percolation was incompatible with resolving the cosmological problems tackled by inflation. In new inflation (and in all subsequent `newer' inflationary models), a graceful exit is a built-in feature. Indeed, a graceful exit is possible regardless of whether inflation takes place at small field values rolling toward large field values (hilltop inflation), or at large field values rolling to the origin in field space (plateau models). The evolution of the field driving inflation, the inflaton,\footnote{The name for this field was first used in \cite{nos} when it was coined by Dimitri Nanopoulos when asked what this field was if we divorce its identity from the SU(5) Higgs adjoint which was the commonly assumed candidate for the field driving inflation.} is governed by its equation of motion and begins with a period of slow-roll characterized by exponential expansion, $a \sim e^{H_I t}$, where $a$ is the cosmological scale factor and $H_I$ is a nearly constant
Hubble parameter during inflation. Slow-roll naturally ends with an oscillatory phase where the inflaton oscillates about a minimum, and a matter-dominated phase, $a \sim t^{2/3}$ begins.

However, at some point, inflaton oscillations must give way
to a radiation-dominated Universe. This transition is most easily accomplished through the decay of the inflaton \cite{dg,nos}, leading to a period of reheating. Provided that the inflaton decay products scatter and thermalize \cite{therm}, the standard radiation-dominated era of the early Universe is produced. In this review, dedicated to the accomplishments of Professor Dimitri Nanopoulos, we revisit the possibilities for inflaton decay in Starobinsky-like models, and explore these same models when derived from no-scale supergravity \cite{no-scale}.

The Starobinsky model \cite{Staro} predates the original models of inflation based on a GUT phase transition, and was an attempt to
find a solution for a singularity-free Universe. The model is based 
on an $R + R^2$ theory of gravity, which is conformally equivalent to a theory that combines Einstein gravity with a scalar field, potentially serving as the inflaton \cite{WhittStelle}. 
Furthermore, the first calculation of a nearly flat scale-independent density fluctuation spectrum was done in this context \cite{MC}.

Shortly after the advent of the new inflationary model, 
it was recognized that supersymmetry may play an important role \cite{enot} in preventing an inflationary hierarchy problem, which requires the inflaton mass to be $\ll M_P$, where $M_P= 2.4 \times 10^{18}$~GeV is the reduced Planck mass. Supergravity is the extension of supersymmetry that incorporates gravitational interactions and it is natural to construct inflationary models consistent with supergravity. However, minimal supergravity models are plagued with the so-called $\eta$ problem \cite{eta}, in which scalars tend to obtain large (Hubble scale) masses, preventing
an extended period of inflation. 
This problem is alleviated \cite{GMO} in no-scale supergravity models \cite{no-scale}, formulated on a maximally symmetric field-space manifold and have been derived as the effective low energy theory from string theory \cite{Witten}.
Of particular interest here is the class of inflationary models
based on no-scale supergravity, which yield a model related to the Starobinsky model \cite{eno6,eno7,enov1}. 

In this contribution, we will focus on inflaton decay in Starobinsky-like models and related models based on no-scale supergravity. Some of these results may be applicable to $T$-model attractors \cite{Kallosh:2013hoa} as well. We begin by reviewing the Starobinsky model, augmented with a matter sector that, for simplicity, consists only of the Standard Model Higgs boson (or the supersymmetric Higgs, as the case may be). We also consider the possibility that the Higgs boson is coupled to curvature (with a coupling $\xi$), where the decay rate of the inflaton vanishes when $\xi = 1/6$. In this case, the dominant decay mechanism comes from the coupling to $T^\mu_\mu$ through the conformal anomaly \cite{Gorbunov:2012ns}. We also consider Starobinsky-like models derived in no-scale supergravity and examine the prospects for decay in these models. In general, inflaton decays in no-scale models are highly suppressed \cite{EKOTY,EGNO4}. In the absence of a direct (superpotential) coupling of the inflaton to the Standard Model, decay to gauge bosons through the gauge kinetic function may be the dominant channel for inflaton decay. For a review on building inflationary models in no-scale supergravity, see Ref.~\cite{building}.

In what follows, we begin with the Starobinsky model as a model of inflation and compare the inflationary observables---the tilt in the perturbation spectrum and the tensor-to-scalar ratio---to CMB results. We consider several 
possibilities for coupling the Standard Model to the $R+R^2$ theory and calculate the inflaton decay rate. Even in the Einstein frame, fields are generally not canonically normalized and we discuss the procedure for determining canonical field coordinates. We also discuss inflaton decay through the conformal anomaly. In Section \ref{sec:no-scale}, we discuss the derivation of a Starobinsky-like model in the context of no-scale supergravity.  We also review the possibilities for inflaton decay in this context. In this case as well, fields are generally non-canonical, and we briefly review the transformation to canonical field coordinates. Finally, in Section \ref{sec:comp}, we relate the formulations of the inflationary model in the contexts of $R+R^2$ gravity and no-scale supergravity. 
There is a close correspondence between the two when the Higgs is conformally coupled with $\xi = 1/6$. Our conclusions are presented in Section \ref{sec:sum}.

\section{The Starobinsky model and inflaton decays}

\begin{figure}[!ht]
\centerline{\psfig{file=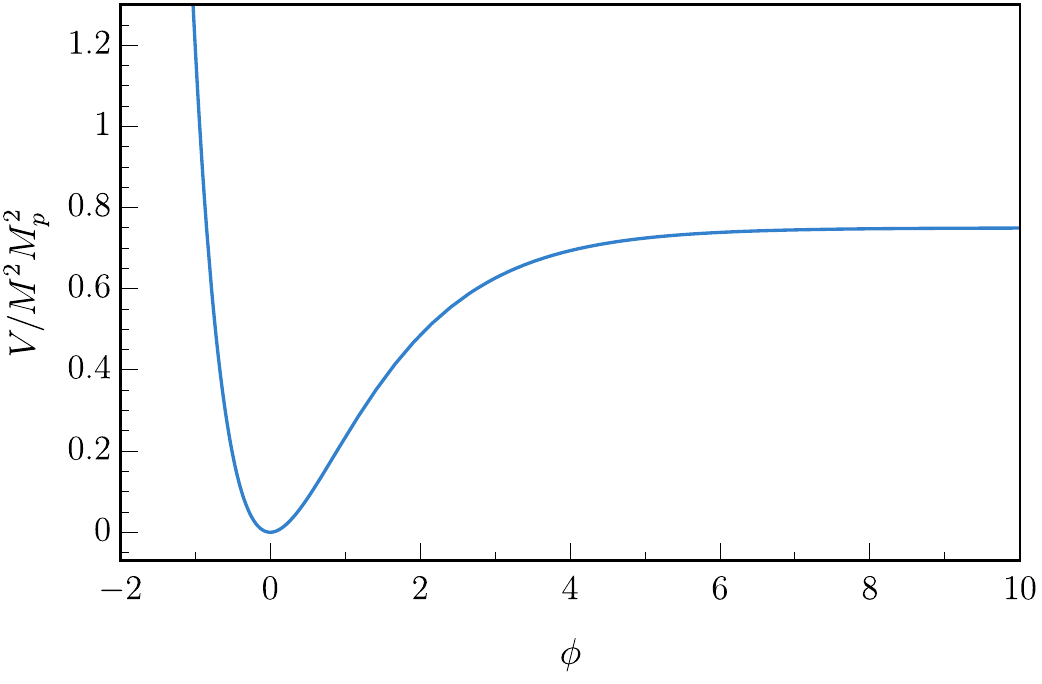,width=8.2cm}}
\caption{\it The scalar potential~(\ref{starpot}) in the Starobinsky model of inflation~\cite{Staro}. \label{fig:staro}}
\end{figure}

The Starobinsky model of inflation~\cite{Staro} can be
formulated by adding an $R^2$ term to the conventional linear Einstein-Hilbert  action:
\begin{equation}
{\cal A} \; = \;  \int d^4x \sqrt{-g} \left(- \frac{1}{2} R + \frac{\alpha}{2} R^2 \right) \, , 
\label{EH}
\end{equation}
where $\alpha$ is a constant. Here, we work in units where $M_P = 1$.  Upon introducing a Lagrange multiplier $-\alpha R^2 \rightarrow 2\alpha \Phi R + \alpha \Phi^2$, this action may be rewritten in the form
\begin{equation}
{\cal A} \; = \; - \frac{1}{2}\int d^4x \sqrt{-g} \left(R + 2 \alpha \Phi R + \alpha\Phi^2 \right) \, .
\label{phiR2}
\end{equation}
Then, following the conformal transformation \cite{WhittStelle,Kalara:1990ar}
\begin{equation}
{\tilde g}_{\mu \nu}  \; = \; e^{2\Omega} g_{\mu \nu} \; = \;  \left(1 + 2 {\alpha} \Phi \right) g_{\mu \nu} \, ,\label{conftransf}
\end{equation}
we can rewrite the action in the Einstein frame as
\begin{equation}
{\cal A} \; = \; - \frac{1}{2}  \int d^4x \sqrt{- {\tilde g}} \left[ {\tilde R} - \frac{6 {\alpha}^2}{(1 + 2 {\alpha}  \Phi)^2} \left(\partial^\mu \Phi \partial_\mu \Phi -  \frac{\Phi^2}{6 {\alpha}} \right) \right] \, .
\label{almostStaro}
\end{equation}
Note the appearance of an additional scalar degree of freedom,
with a non-canonical kinetic term and potential $V(\Phi)$. 
By making a field redefinition, $\phi \equiv \sqrt{3/2} \ln \left(1 +  2 { \alpha} \Phi \right)$, Eq.~(\ref{almostStaro}) may be written as:
\begin{equation}
{\cal A} \; = \; - \frac{1}{2} \int d^4x \sqrt{- {\tilde g}} \left[ {\tilde R} - \partial^\mu \phi \partial_\mu \phi + \frac{1}{4 {\alpha}} 
\left(1 - e^{-\sqrt{\frac{2}{3}} \phi }
 \right)^2 \right] \, .
\label{FullStaro}
\end{equation}
When ${\alpha} = 1/6 M^2$, the potential is now seen with the well-known form of the Starobinsky potential,
\beq
V(\phi) \; = \; \frac34 M^2\left(1 - e^{-\sqrt{\frac{2}{3}} \phi }
 \right)^2 \, .
 \label{starpot}
\eeq
This potential is shown in Fig.~\ref{fig:staro}.

As an inflationary model, the potential (\ref{starpot}) leads to observables in excellent agreement with observations by Planck and other CMB experiments \cite{Planck,rlimit,Tristram:2021tvh}. The comparison to experiment is easily made by first computing the slow-roll parameters, $\epsilon$ and $\eta$ for single field inflationary models, 
\begin{equation} 
\epsilon \; \equiv \; \frac{1}{2} M_{P}^2 \left( \frac{V'}{V} \right)^2 ; \; \;  \eta \; \equiv \; M_{P}^2 \left( \frac{V''}{V} \right)   \, ,
\label{epsilon}
\end{equation}
where the prime denotes a derivative with respect to the inflaton field $\phi$.
Of particular interest is the tilt in the spectrum of scalar perturbations, $n_s$ \cite{Planck},
\begin{eqnarray}
 n_s \; & \simeq &\; 1 - 6 \epsilon_* + 2 \eta_*\\
& = &\; 0.965 \pm 0.004 \; (68\%~{\rm CL}) \, ,
\label{ns} 
\end{eqnarray}
and the tensor-to-scalar ratio, $r$ \cite{rlimit},
\beq
r \; 
\simeq  \; 16 \epsilon_* < 0.036 \; (95\%~{\rm CL}) \, . \label{r}
\eeq
The overall inflationary scale, $M$, is set by the amplitude of scalar perturbations, $A_s$ \cite{Planck},
\beq
A_s \; = \; \frac{V_*}{24 \pi^2 \epsilon_* M_{P}^4 } \simeq 2.1 \times 10^{-9} \, . \label{As} 
\eeq
The number of $e$-folds, $N_i$, of inflation between the initial and final values of the inflaton field
is given by the formula
\begin{equation}
N_i \;\equiv\; \ln\left(\frac{a_{\rm{end}}}{a_i}\right) \; = \; \int_{t_i}^{t_{\rm{end}}} H dt \; \simeq \;  - \int^{\phi_{\rm{end}}}_{\phi_i} \frac{1}{\sqrt{2 \epsilon}} \frac{d \phi}{M_P} \, .
\label{efolds}
\end{equation}
The number of $e$-folds between when the pivot scale $k_*$ exits the horizon and the end of inflation is denoted by $N_*$. Inflation ends when $a=a_{\rm end}$ and $\ddot{a} = 0$. Typical values of $N_*$ are in the range $\sim$ 50-60, depending on the mechanism that ends inflation~\cite{LiddleLeach,MRcmb,Planck}.

For the Starobinsky model, it is straightforward to compute the inflationary observable listed above~\cite{eno6}:
\begin{eqnarray}
A_s &  = &  \frac{3 M^2}{8\pi^2} \sinh^4 (\phi/\sqrt{6}) \, , \label{As2} \\
\epsilon & = & \frac13 \csch^2 (\phi/\sqrt{6}) e^{-\sqrt{2/3}\phi} \, ,  \label{eps}\\
\eta & = &  \frac13 \csch^2 (\phi/\sqrt{6}) \left( 2 e^{-\sqrt{2/3}\phi} -1\right) \label{ns} \, .
\end{eqnarray} 
For the Starobinsky potential, $\phi_{\rm end} = \phi(a_{\rm end}) = 0.62 M_P$ \cite{EGNO5,egnov}
and $N_*=55$ for $\phi_* = 5.35M_P$. In this case, we find $M = 1.25 \times 10^{-5} M_{P} \simeq 3 \times 10^{13}~{\rm GeV}$, $n_s = 0.965$, and $r = 0.0035$.

\begin{figure}[!ht]
\centerline{\psfig{file=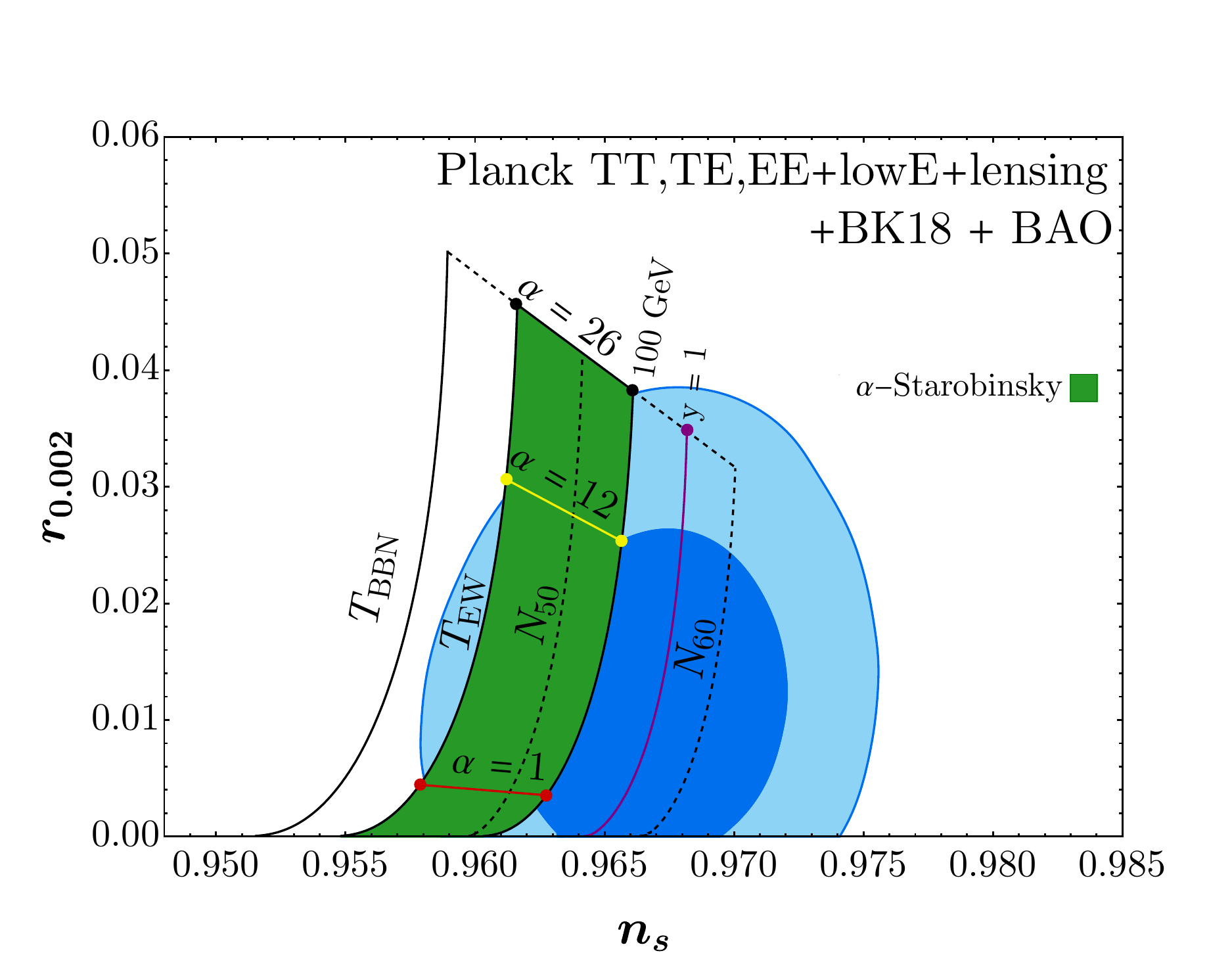,width=10cm}}
\caption{\it Plot of the CMB observables $n_s$ and $r$.  
The blue shadings correspond to the 68\% and 95\%
confidence level regions from Planck data combined with BICEP2/Keck results~\cite{Planck,rlimit}.
The red dots show the predictions of the $\alpha$-Starobinsky potential with $\alpha = 1$ (i.e., the Starobinsky model) for $N_* \sim 46$ (left) and $\sim 52$ (right) corresponding to limits from reheating  100 GeV $< \trh < 10^{10}$~GeV. 
The upper pair of yellow (black) dots are the predictions when $\alpha = 12~(26)$ 
(the largest value of $\alpha$ consistent with 68\% CL~(95\% CL) CMB observations and $\trh \lesssim 10^{10}$~GeV. The vertical lines correspond to the prediction assuming a reheat temperature (or value of $N_*$ as appropriate) of $\trh = 2$ MeV, $\trh = 100$ GeV, $N_* = 50$, $\trh \simeq 10^{10}$~GeV, $\trh \simeq 5 \times 10^{14}$~GeV, and $N_* = 60$.  
\label{fig:plancklimits}}
\end{figure}

We show in Fig.~{\ref{fig:plancklimits}} taken from Ref.~\cite{egnov}, the 68\% and 95\% CL regions of the $(n_s, r)$ plane allowed by the Planck and Keck/BICEP2 data~\cite{Planck,rlimit}. The (nearly) horizontal line (labeled $\alpha =1$)\footnote{The parameter $\alpha$ here refers to a generalization of the Starobinsky model where the exponent in Eq.~(\ref{starpot}) becomes $-\sqrt{\frac{2}{3\alpha}} \phi$ \cite{eno7,KLR,ENOV3}. } corresponds to the predictions of the Starobinsky model~(\ref{starpot}) for $N_* \sim 46$ to $\sim 52$ e-folds, corresponding to limits from reheating (see below) 100 GeV $< \trh < 10^{10}$~GeV, the latter corresponding to the upper limit on $\trh$ from the limit on gravitino production and the relic density of a 100 GeV lightest supersymmetric particle. The other nearly vertical curves correspond to (from left to right), $\trh = 2$~MeV (solid), $N_* = 50$ (dotted), $\trh \simeq 5 \times 10^{14}$~GeV, obtained when a perturbative decay coupling is $\mathcal{O}(1)$ (with $\Gamma_{\phi} = \frac{y^2}{8 \pi} m_{\phi}$) (solid), and $N_* = 60$ (dotted).   
For the $\alpha$-Starobinsky model~\cite{eno7,KLR,ENOV3}, we shade the region in green that respects the constraint $T_{\rm reh} > T_{\rm EW}$ and the relic density constraint of $m_{\rm LSP} \simeq 100$ GeV. For more details, see~\cite{egnov}.

As described above, the $\alpha$-Starobinsky model will inflate and produce density fluctuation in agreement with the experiment. 
When $\phi < \phi_{\rm end}$, the inflaton begins a period of oscillations, and the energy density in these oscillations
scales as $a^{-3}$, typical of a matter-dominated universe. 
These oscillations continue until inflaton decays occur sufficiently fast to produce radiation.  How this occurs depends on how the Standard Model is introduced in the context of inflationary theory and how the inflaton couples to the Standard Model (SM).

We first consider a simple model by adding a complex scalar field $H$ to the $R+R^2$ action:
\begin{equation}
{\cal A}_1 \; = \;  \int d^4x \sqrt{-g} \left(- \frac{1}{2} R + \frac{\alpha}{2} R^2+\partial^\mu H^* \partial _\mu H-V(H) \right) \, . 
\label{model1}
\end{equation}
After performing the same conformal transformation \eqref{conftransf} and  the field redefinition $\phi \equiv \sqrt{3/2} \ln \left(1 +  2 { \alpha} \Phi \right)$, the above action can be rewritten as
\begin{equation}
    \begin{aligned}
        \label{eq:acteinst}
        {\cal A}_1 \; = \; - \frac{1}{2} \int d^4x \sqrt{- {\tilde g}} \biggl[& {\tilde R} - \partial^\mu \phi \partial_\mu \phi + \frac{1}{4 {\alpha}} 
\left(1 - e^{-\sqrt{\frac{2}{3}} \phi }
 \right)^2  \\& -2e^{-\sqrt{\frac{2}{3}}\phi}\partial^\mu H^* \partial _\mu H+2e^{-2\sqrt{\frac{2}{3}}\phi}V(H)  \biggr] \, .
\end{aligned}
\end{equation}
Expanding the second line around the minimum, $\left<\phi\right>=0$, we obtain a canonical kinetic term for the complex scalar, $\partial^\mu H^* \partial _\mu H$, as well as the couplings of the inflaton  to $H$: 
 \begin{equation}
    {\cal A}_1 \; \supset \;  \int d^4x \sqrt{- {\tilde g}}\left[-\sqrt{\frac{2}{3}}\phi \partial^\mu H^* \partial _\mu H+2\sqrt{\frac{2}{3}}\phi V(H)\right]\,.
    \label{a1coupl}
 \end{equation}
The dominant contribution to the decay $\phi \to HH$ comes from the first term (the 2nd term gives a contribution which is suppressed by the Higgs mass) and the rate is given by
\begin{align}
	\Gamma(\phi \to HH) = \frac{N_H}{192\pi}\frac{M^3}{M_P^2}\,,
	\label{eq:decay_R2_A1}
\end{align}
where $N_H = 4$ is
the number of the real scalar degrees of freedom of the SM Higgs doublet.

It is often useful to perform a field redefinition to obtain canonical fields in a particular background using the Riemann normal coordinates (RNC). The term $\phi \partial^\mu H^* \partial _\mu H$ can be eliminated and the coupling of the inflaton to Higgs is purely a potential interaction. The RNC fields can be obtained quite generally. Consider a Lagrangian of the form
\begin{equation}
   \mathcal{L} \; = \; \frac{1}{2}g_{IJ}(\Phi) \partial_{\mu} \Phi^I \partial^{\mu} \Phi^J - V(\Phi) \, ,
\end{equation}
where $g_{IJ}(\Phi)$ is a symmetric metric and $V(\Phi)$ is the potential. Here we use the Latin indices $I,J, ..., $ for the flavor eigenbasis. To compute the interactions around the minimum, we introduce the field expansion $\Phi^I = v^I + \phi^I$, where $v^I$ is the vacuum expectation value and $\phi^I$ is the field fluctuation. Following the discussions in~\cite{Cheung:2021yog,swamp1, swamp2}, the symmetrized covariant derivatives of the potential can be expressed as
\begin{equation}
    \label{eq:covder}
   V_{I_1 \ldots I_k}(v) \; = \; \left.\nabla_{\left(I_1 \ldots\right.} \nabla_{\left.I_k\right)} V(\Phi)\right|_v \, , \quad \text { with }\left.\quad \partial_I V(\Phi)\right|_v=0 \, ,
\end{equation}
where the covariant derivatives are symmetrized and include all permutations multiplied by a factor of symmetry $1/n!$. Here we take the derivatives with respect to field $\Phi^I$, evaluated at their VEV $v^I$. The Christoffel symbols are given by
\begin{equation}
    \label{eq:christoffel}
   \Gamma^{I}_{JK} \; = \; \frac{1}{2} g^{IL} \left(g_{KL,J} + g_{LJ,K} - g_{JK, L}\right) \, .
\end{equation}
Using these definitions, we find that the canonically normalized mass matrix at the minimum is given by $V_{IJ}(v) = \partial_I \partial_J V(\Phi)|_v$. To canonically normalize the fields and transform from the flavor to mass eigenbasis, we introduce a tetrad that flattens the metric
\begin{equation}
    g_{IJ}(v) e_i^I(v) e_j^J(v) \; = \; \delta_{ij} \, .
\end{equation}
Here, the lowercase letters correspond to the mass eigenbasis, and the indices may be raised or lower using the Kronecker delta $\delta_{ij}$ or $\delta^{ij}$. The inverse tetrad can be computed using the expression $e^i_{I} = \delta^{ij}e_j^{~J}g_{JI}$.

In Riemann normal coordinates, there are no cubic derivative interactions, which significantly simplifies the computation. Thus, in this case, the cubic interaction can arise only from the potential terms. From the action~(\ref{eq:acteinst}), we find that the field metric of the kinetic terms $\frac{1}{2}g_{IJ}(\Phi) \partial_{\mu}\Phi^I \partial^{\mu} \Phi^J$ can be expressed as
\begin{equation}
    g_{IJ} \; = \; {\rm{diag}}\left(1 \,, 2 e^{-\sqrt{\frac{2}{3}}\phi} \,, 2e^{-\sqrt{\frac{2}{3}}\phi}\right) \, ,
\end{equation}
where $\Phi = (\phi,\, \mathrm{Re} \, H,\, \mathrm{Im} \, H)$. At the minimum, the VEVs of these fields are $v = (0, 0, 0)$. We find that the tetrad at the minimum is given by
\begin{equation}
    e_i^I(v) = {\rm diag}\left(1, \frac{1}{\sqrt{2}}, \frac{1}{\sqrt{2}} \right) \, .
\end{equation}
Finally, we compute the three-particle scattering amplitude using Eq.~(\ref{eq:covder}) that includes a symmetry factor for two identical final states in the amplitude. To compute the scattering amplitude in the mass eigenbasis, we use
\begin{equation}
    V_{IJK}(v) e_{i}^I(v) e_j^J(v) e_k^K(v) \; = \; V_{ijk}(v) \, ,
\end{equation}
and find
\begin{equation}
    V_{\phi,\, \mathrm{Re} \, H ,\, \mathrm{Re} \, H} \; = \; V_{\phi,\, \mathrm{Im} \, H,\, \mathrm{Im} \, H}  \; = \; -\frac{M^2}{\sqrt{6}} \, .
\end{equation}
These couplings lead to the decay rate~(\ref{eq:decay_R2_A1}). For more details on how to use the RNC to compute the scattering amplitudes, see Refs.~\cite{Cheung:2021yog,swamp1, swamp2}.

Another alternative description of the model entails a conformal transform into 
the conformal frame
where all the scalar fields conformally couple to the gravity.
This is achieved by the conformal transformation
\begin{align}
	g_{c\mu\nu} = e^{2{\Omega}_c}g_{\mu\nu}\,,
\end{align}
with the inflaton $\phi$ and the conformal factor satisfying
\begin{align}
	\phi_c = \sqrt{6}\left(e^{-{\Omega}_c}-1\right)\,,
	\quad
	1+2\alpha \Phi = e^{2{\Omega}_c}\left(1-\frac{\phi_c^2+2\vert H_c\vert^2}{6}\right)\,,
	\quad
	H = e^{{\Omega}_c}H_c\,.
\end{align}
Then the above action can be rewritten as~\cite{Ema:2020zvg}
\begin{align}
	\mathcal{A}_1 = \int d^4x \sqrt{-g_c}
	&\left[-\frac{1}{2}R_c\left(1-\frac{\phi_c^2 + 2\vert H_c\vert^2}{6}\right) 
	+ \frac{1}{2}\partial^\mu \phi_c \partial_\mu \phi_c + \partial^\mu H_c^* \partial_\mu H_c
	\right. \nonumber \\ &\left.
	- \left(1+\frac{\phi_c}{\sqrt{6}}\right)^4 V\left(\frac{H_c}{1+\phi_c/\sqrt{6}}\right)
	- \frac{3M^2}{4}\left(\sqrt{\frac{2}{3}}\phi_c + \frac{\phi_c^2}{3} + \frac{\vert H_c\vert^2}{3}\right)^2
	\right]\,.
	\label{eq:action_R2_A1}
\end{align}
In this frame, the scalar fields are canonical 
and conformally coupled to gravity without any kinetic mixing,
and the decay is induced by the cubic interaction $\phi_c \vert H_c \vert^2$ from the potential.
Focusing on the Standard Model Higgs with $V(H) = m^2 \vert H\vert^2 + \lambda \vert H \vert^4$,
the decay is predominantly induced by the last term in Eq.~\eqref{eq:action_R2_A1},
and we recover the decay rate given in Eq.~\eqref{eq:decay_R2_A1},
which is expected due to frame independence.

One may also introduce a non-minimal coupling to gravity $\xi R H^*H$:
\begin{equation}
{\cal A}_2 \; = \;  \int d^4x \sqrt{-g} \left(- \frac{1}{2} R + \frac{\alpha}{2} R^2+\partial^\mu H^* \partial _\mu H+\xi R H^*H -V(H)\right) \, .
\label{model2}
\end{equation}
The conformal transformation in this case is
\begin{equation}
{\tilde g}_{\mu \nu}  \; = \; e^{2\Omega} g_{\mu \nu} \; = \; \left(1 + 2 {\alpha} \Phi - 2 \xi |H|^2 \right) g_{\mu \nu}  \, .
\label{newtilde}
\end{equation}
In the Einstein frame, this is equivalent to
\begin{equation}
\begin{aligned}
{\cal A}_2 \; = \; - \frac{1}{2} \int d^4x \sqrt{- {\tilde g}} \biggl[& {\tilde R} - \partial^\mu \phi \partial_\mu \phi + \frac{1}{4 {\alpha}} 
\left(1 - e^{-\sqrt{\frac{2}{3}} \phi }+2\xi e^{-\sqrt{\frac{2}{3}} \phi}H^*H 
 \right)^2  \\& -2e^{-\sqrt{\frac{2}{3}}\phi}\partial^\mu H^* \partial _\mu H+2e^{-2\sqrt{\frac{2}{3}}\phi}V(H)  \biggr] \, , \end{aligned}
\end{equation}
where now $\phi = \sqrt{3/2} \ln (1 + 2 \alpha \Phi -2 \xi |H|^2)$. 

To compute the decay rate in this frame, one can again expand the couplings around the minimum, which gives rise to the following dominant trilinear couplings:
 \begin{equation}
    {\cal A}_2 \; \supset \;  \int d^4x \sqrt{- {\tilde g}}\left[-\frac{1}{2\alpha}\xi \sqrt{\frac{2}{3}}\phi|H|^2 -\frac{2}{\sqrt{6}}\phi\partial^\mu H^*\partial_\mu H\right]\,.
    \label{a2coupl}
 \end{equation}
The rate is then\footnote{See also \cite{Watanabe:2006ku} for discussions of inflaton decay in these models.}
\begin{align}
	\Gamma(\phi \to HH) = \frac{N_H(1-6\xi)^2}{192\pi}\frac{M^3}{M_P^2}\,.\label{ratea2ein}
\end{align}
Alternatively, we may
move to the conformal frame by
\begin{align}
	g_{c\mu\nu} = e^{2{\Omega}_c}g_{\mu\nu}\,,
\end{align}
with the inflaton $\phi$ and the conformal factor satisfying
\begin{align}
 	\phi_c = \sqrt{6}\left(e^{-{\Omega}_c}-1\right)\,,
	\quad
	1+2\alpha \Phi - 2\xi \vert H\vert^2 = e^{2{\Omega}_c}\left(1-\frac{\phi_c^2+2\vert H_c\vert^2}{6}\right)\,,
	\quad
	H = e^{{\Omega}_c}H_c\,,
\end{align}
and we obtain~\cite{Ema:2020zvg}
\begin{align}
	 \mathcal{A}_2 = \int d^4x \sqrt{-g_c}
	&\left[-\frac{1}{2}R_c\left(1-\frac{\phi_c^2 + 2\vert H_c\vert^2}{6}\right) 
	+ \frac{1}{2}\partial^\mu \phi_c \partial_\mu \phi_c + \partial^\mu H_c^* \partial_\mu H_c
	\right. \nonumber \\ &\left.
	- \left(1+\frac{\phi_c}{\sqrt{6}}\right)^4 V\left(\frac{H_c}{1+\phi_c/\sqrt{6}}\right)
	- \frac{3M^2}{4}\left(\sqrt{\frac{2}{3}}\phi_c + \frac{\phi_c^2}{3} + \frac{1-6\xi}{3}\vert H_c\vert^2\right)^2
	\right]\,.
\end{align}
This form of action makes it manifest that 
the interaction between the inflaton and the Higgs is controlled by the combination $1-6\xi$.
The decay is again induced by the cubic interaction $\phi_c \vert H_c \vert^2$ from the last term in the potential,\footnote{
	Even if we assume that the Higgs mass is of order the electroweak scale after integrating out the inflaton,
	the inflaton coupling to the Higgs induces an additional Higgs mass term above the inflaton mass scale~\cite{Ema:2020evi}.
	However, as long as $\xi$ is small, the generated mass is suppressed compared to $M$, therefore it has a negligible contribution
	to the decay rate.
}
and not surprisingly 
the rate is given by Eq.~(\ref{ratea2ein}). 
Note that a very large negative value for $\xi$ forces the Higgs to participate in the inflationary dynamics,
making it the so-called Higgs-$R^2$ inflation model~\cite{Wang:2017fuy}. 
In this case, the (p)reheating dynamics can be non-perturbative~\cite{He:2018mgb},
and we do not consider this regime in the following.

The above formula shows that, if the Higgs conformally couples to gravity, with $\xi = 1/6$,
the inflaton does not decay into the Higgs.
In this case, the inflaton predominantly decays into Standard Model gauge bosons
through the trace anomaly~\cite{Gorbunov:2012ns}.
In general, the coupling between the inflaton and Standard Model particles in the Starobinsky model is expressed to the lowest order as
\begin{align}
	\mathcal{L} = \frac{1}{\sqrt{6}} \phi {T^\mu}_\mu\,,
 \label{OT}
\end{align}
where ${T^\mu}_\mu$ is the trace of the stress-energy tensor.
One can see that Eqs.~\eqref{a1coupl} and~\eqref{a2coupl} indeed take this form.
If the matter sector is (classically) conformal, 
and assuming the absence of direct coupling between the gauge sector 
and the gravity beyond the minimal one in the original frame,
${T^\mu}_\mu$ is predominantly sourced by the trace anomaly.
The origin of the trace anomaly depends on the regularization scheme 
(although the final result itself does not)~\cite{Kamada:2019pmx}.
For instance, in dimensional regularization, the inflaton couples to
the gauge bosons as $\epsilon \phi F_{\mu\nu}F^{\mu\nu}$, with the spacetime dimension $d = 4-2\epsilon$, 
after the conformal transformation.
This factor of $\epsilon$ is compensated by the $1/\epsilon$ pole arising from the wavefunction renormalization
(and self-interactions in the non-abelian case),
resulting in a finite term in the limit $\epsilon \to 0$ proportional to
$\beta \phi F_{\mu\nu}F^{\mu\nu}$, where $\beta$ is the beta function
counting \emph{both} heavy and light degrees of freedom.
In addition, the inflaton directly couples to heavy degrees of freedom through their mass terms.
After integrating out heavy degrees of freedom, the contribution from the mass term cancels 
with the one from the wavefunction renormalization,
leaving only light degrees of freedom, as demonstrated in~\cite{Kamada:2019pmx}.
As a result, the trace anomaly is given by
\begin{align}
	\left.{T^\mu}_\mu\right\vert_\mathrm{anom} = \sum_{i,a} \frac{\beta_i}{4\alpha_i} F^{a}_{i\mu\nu} F_i^{a \mu\nu}\,,
 \label{anomT}
\end{align}
where $i$ runs over all the different gauge groups, $a$ runs over the generators of each gauge group,
and $\alpha_i$ is the fine structure constant of the gauge group $i$.
The beta function is given by
\begin{align}
	\beta_i = \frac{b_i \alpha_i^2}{2\pi}\,,
\end{align}
where $b_i$ counts only the number of light degrees of freedom as explained above,
and $b_1 = 41/10~(33/5)$ for U(1),\footnote{
	Note that this corresponds to the beta function for $g_1 = \sqrt{5/3}\,g'$.
} $b_2 = -19/6~(1)$ for SU(2), and $b_3 = -7~(-3)$ for SU(3) for the Standard Model 
(the Minimal Supersymmetric Standard Model), respectively.
The decay rate of the inflaton to the gauge bosons is then given by
\begin{align}
	\Gamma(\phi \to AA) = \sum_i \frac{N_{A_i} b_i^2 \alpha_i^2}{768\pi^3}\frac{M^3}{M_P^2}\,,
 \label{anomD}
\end{align}
where $N_{A_i}$ is the number of the gauge bosons in the gauge group $i$,
and we again explicitly include the Planck scale.
To evaluate this expression, we use the couplings at the scale of $M/2$ after the renormalization group running.
In the case of the Standard Model, by running the couplings up to two-loop with \texttt{SARAH}~\cite{Staub:2013tta} 
and taking the input values at the electroweak scale following~\cite{Buttazzo:2013uya}, 
we obtain $\alpha_1(M/2) \simeq \alpha_2(M/2) \simeq 0.024$ and $\alpha_3(M/2) \simeq 0.027$,
where $M \simeq 3\times 10^{13}\,\mathrm{GeV}$.

There are of course numerous other ways to introduce the SM Lagrangian. For example, one can introduce it directly in the Einstein frame
leading to an action similar to the one in Eq.~(\ref{eq:acteinst}) without the coupling of the inflaton to the Higgs kinetic and potential terms. In this case, there is no coupling of the inflaton to matter, though to achieve this starting from the Jordan frame is quite contrived. Hence 
we restrict ourselves to the above simple possibilities.

If there is a non-negligible decay width for the inflaton, 
its decays will start populating the radiation bath
and initially there will be a rapid increase in radiation temperature followed by a slow redshift with $\rho_{r} \propto a^{-3/2}$ \cite{Giudice:2000ex,GKMO1,GKMO2}. 
We define the reheating temperature corresponding to $\rho_r(\arh) = \rho_\phi(\arh)$, or
\begin{align}
	\frac{g_{\rm RH} \pi^2}{30} \trh^4 = \frac{12}{25} \left(\Gamma_\phi M_P \right)^2 \, ,
 \label{deftrh}
\end{align}
where $g_{\rm RH}$ is the number of degrees of reheating at $\trh$. Subsequently, the radiation will redshift normally as $\rho_r \propto a^{-4}$.  In the SM, $g_{\rm RH} = 427/4$ and for $M =  3 \times 10^{13}$~GeV,
we find
\begin{align}
	\trh = \mathrm{max}\left[2.9\times 10^{9}\,\mathrm{GeV}\times \vert 1 - 6\xi \vert,~
	1.3\times 10^{8}\,\mathrm{GeV}\right]\,,
\end{align}
where the former comes from $\phi \to HH$ while the latter from $\phi \to AA$.
Thus, in the Starobinsky model, there is always a significant source of reheating. 

Before concluding this section, we note that many of the above arguments can be applied to $T$-model attractors \cite{Kallosh:2013hoa}.
Although originally formulated using an action with an O(1,1) symmetry, these models can also be simply derived from a Jordan frame with
an action \cite{Kallosh:2013maa}
\begin{equation}
{\cal A} \; = \;  \int d^4x \sqrt{-g} \left(- \frac{1}{2} R f(\phi) + \frac12 \partial^\mu \phi \partial_\mu \phi - V(f) \right) \, , 
\label{EHT}
\end{equation}
with $f(\phi) = 1 - \phi^2/6$.
After a transformation to the Einstein frame
we have 
\begin{equation}
{\cal A} \; = \;  \int d^4x \sqrt{-g} \left(- \frac{1}{2} R + \frac{1}{2f^2} \partial^\mu \phi \partial_\mu \phi - \frac{1}{f^2} V(f) \right) \, .
\label{EHTE}
\end{equation}
Taking $V(f) = \lambda f^2 \phi^k$ and making the field redefinition $\phi = \sqrt{6} \tanh(\chi/\sqrt{6})$, we obtain
\begin{equation}
{\cal A} \; = \;  \int d^4x \sqrt{-g} \left(- \frac{1}{2} R + \frac{1}{2} \partial^\mu \chi \partial_\mu \chi - \lambda (\sqrt{6} \tanh(\chi/\sqrt{6}))^k \right) \, .
\label{EHTE2}
\end{equation}
Possible decay modes of the inflaton $\chi$,
can be obtained through the induced couplings to the Higgs bosons (once introduced) or through the gauge anomaly.

\section{No-scale inflationary models and inflaton decay}
\label{sec:no-scale}

The bosonic sector of an $\mathcal{N} = 1$ supergravity theory is specified by a K\"ahler potential, $K(\phi^i,\phi_j^*)$, which determines the field-space geometry of the chiral scalar fields in the theory, a holomorphic function of these fields, $W(\phi^i)$, which determines the interactions between these fields and their fermionic partners, and a gauge kinetic function, $f_{\alpha\beta}(\phi^i)$. Taken together, the bosonic Lagrangian
can be written as 
\beq
{\mathcal L} = - \frac12 R + K^j_i \partial_\mu \phi^i \partial^\mu \phi^*_j - V - \frac14 {\rm Re} (f_{\alpha\beta}) F^\alpha_{\mu\nu} F^{\beta \mu\nu} -\frac{1}{4} {\rm Im} (f_{\alpha\beta}) F^\alpha_{\mu\nu} \tilde{F}^{\beta\mu\nu} \, ,
\label{Lkin3J}
\eeq
where the first term is the minimal Einstein-Hilbert
term of general relativity and in the second term $K^j_i \equiv \partial^2 K/\partial \phi^i \partial \phi_j^*$ is the field-space metric.
The effective scalar potential,
\begin{equation}
\label{effpot}
V \; = \; e^{G} \left[G_i \left({G^{-1}}\right)^i_j G^j -3 \right] \, ,
\end{equation}
where 
\begin{equation}
G \; \equiv \;  K + \ln~|W|^2 \, ,
\end{equation}
and
$G_i \equiv \partial{G}/{\partial \phi^i}$, $G^j \equiv \partial{G}/{\partial\phi_j^*}$, and $\left(G^{-1}\right)^i_j$ is the inverse of the matrix of second
derivatives of $G$. In addition, 
there are also $D$-term contributions for gauge non-singlet
chiral fields. For a review of local supersymmetry, see Ref.~\cite{susy}.
Minimal supergravity (mSUGRA) is characterized by a K{\" a}hler potential
of the form 
\beq
K \; = \; \phi^i \phi_i^* \, ,
\label{Kmin}
\eeq 
in which case the effective potential (\ref{effpot}) can be written in the form
\beq
V(\phi^i,\phi_j^*) \; = \; e^{\phi^i {\phi^*_i}}  \left[ |W_i + \phi_i^* W
|^2 - 3|W|^2
\right] \,, 
\label{sgpotJ}
\eeq
where $W_i \equiv \partial{W}/{\partial\phi^i}$. 

Unlike the scalar potential in globally supersymmetric models, where  $V = |W_i|^2$, 
the minimal supergravity potential is {\it not} positive
semi-definitive. Indeed, the negative term $\propto |W|^2$ in
(\ref{sgpotJ}) generates in general minima with 
$V \sim - {\cal O}(m_{3/2}^2 M_P^2)$ \cite{Ovrut:1983my}, where $m_{3/2}^2 = e^G$ is the gravitino mass. In addition, as noted earlier, it is difficult to generate flat directions suitable for inflation, as scalar fields tend to pick up order $H^2$ masses \cite{eta}.

In contrast, such difficulties are absent in no-scale supergravity. 
The simplest such (single field) theory is defined by~\cite{no-scale,EKN1}
\begin{equation}
        K \; = \; -3 \ln (T + T^*) \, .
\label{CFKN}
\end{equation}
This describes a maximally symmetric field space with constant curvature ${\mathcal R} = 2/3$ and for $W = 0$,
$V = 0$. The theory is generalized by adding additional scalar fields
with~\cite{ELNT,EKN}
\beq
\label{v0}
K \; = \; -3 \ln (T + T^* - |\phi_i|^2/3) \, .
\eeq
This K\"ahler potential still describes a maximally symmetric field-space with curvature ${\mathcal R} = (N+1)/3$, where $N-1$ is the number of `matter' fields, $\phi^i$. 
In this case, the Lagrangian becomes
\begin{eqnarray}
{\mathcal L}  & = & -\frac12R + \frac{1}{12} (\partial_\mu K)^2 + e^{K/3} |\partial_\mu \phi^i|^2 \nonumber \\
&& +\frac34 e^{2K/3} |\partial_\mu (T- T^*) - \frac13 (\phi_i^{*}\partial_{\mu}\phi^i-\phi^i\partial_{\mu}\phi^{*}_i)|^2  - V \, ,
\label{LmanyJ}
\end{eqnarray}
where the effective scalar potential can be written as 
 \beq
 V \; = \; e^{\frac{2}{3}K} {\hat V} \; = \; \frac{\hat V}{\left((T + {T^*})- \frac{1}{3} |\phi^i |^2 \right)^2} \, ,
 \label{VJ}
 \eeq
 with
  \beq
{\hat V} \; \equiv \; \ \left|  W_{i} \right|^2  +\frac{1}{3} (T+T^*) |W_T|^2 +
\frac{1}{3} \left(W_T (\phi_i^* W^{*i} - 3 W^*) + {\rm h.c.}  \right) \, .
\label{effVJ}
\end{equation}
When $W_T = 0$, the potential takes a form related to that in global supersymmetry,
with a proportionality factor of $e^{2K/3}$, where $K$ is
the canonically-redefined modulus. Large mass terms are not generated~\cite{deln} in this case, and
the $\eta$-problem is avoided~\cite{GMO}. 

Interestingly, no-scale supergravity can be used as the framework to construct models of inflation~ \cite{GL,KQ,EENOS,otherns}. 
In particular, by adopting a very simple (Wess-Zumino) form for the superpotential of a single matter field, $\phi$,\footnote{To construct a Starobinsky-like model, at least two fields, $T$ and $\phi$ are needed \cite{eno7}.} of the form \cite{eno6}
\beq
W_{\rm I} = M \left( \frac12 \phi^2 - \frac{1}{3\sqrt{3}} \phi^3 \right) \, ,
\label{WWZ}
\eeq
the Starobinsky model potential (\ref{starpot}) is obtained once a field redefinition is made to a canonically normalized field $\chi$, 
\beq
\chi \; \equiv \; \sqrt{3} \tanh^{-1} \left( \frac{\phi}{\sqrt{3}} \right) \, ,
\eeq
and $T$ is stabilized with a vacuum expectation value assumed here to be $\langle T \rangle = \frac12$.
Decomposing $\chi$ into its real and imaginary parts:
$\chi = (x + iy)/\sqrt{2}$, the potential is minimized for $y=0$, and in the real direction we obtain 
the Starobinsky potential with the identification of $\phi$ in Eq.~(\ref{starpot}) with $x$ here.

Note that the choice of superpotential in Eq.~(\ref{WWZ}) is not unique. 
There is another well-studied example, defined by \cite{Cecotti, FeKR,EGNO2,others,EGNO3}
\beq
W_{\rm I} = \sqrt{3} M \phi \left(T - \frac{1}{2} \right) \, .
\label{cec}
\eeq
In this case, the Starobinsky model potential (\ref{starpot}) is obtained once a field redefinition is made to a canonically normalized field, 
\begin{equation}\label{Tgen}
T = \frac{1}{2} e^{\pm \sqrt{\frac{2}{3}} t} \, ,
\end{equation}
where $t$ is real  ($t$ here plays the role of the inflaton $\phi$ in Eq.~(\ref{starpot}))
and $\langle \phi \rangle = 0$. Indeed there are multiple classes of such models all related by the underlying  SU(2,1)/SU(2)$\times$U(1) no-scale symmetry \cite{enov1}.

Inflation in all of these models is indistinguishable from the original Starobinsky model, and the inflationary observables
are the same as those given in Eqs.~(\ref{As})-(\ref{ns}).
However, as in the discussion of the previous section,
the matter sector must also be considered in order to achieve reheating and a radiation-dominated universe. In this case, there are a few possibilities. If we are considering only the possibility 
for decays to the Higgs boson, the superpotential must include
a $\mu$-term (in the minimal supersymmetric Standard Model (MSSM)),
\beq
W_{\rm SM} = \mu H_u H_d \, .
\eeq
The Higgs kinetic terms may arise from the K\"ahler potential either as untwisted fields
\beq
\label{v1}
K \; = \; -3 \ln (T + T^* - |\phi|^2/3 -|H_u|^2/3 - |H_d^2|/3) \, ,
\eeq
or as twisted fields 
\beq
\label{v2}
K \; = \; -3 \ln (T + T^* - |\phi|^2/3 ) + |H_u|^2 + |H_d^2| \, .
\eeq

For the case of untwisted fields and the inflaton is $\phi$ (as opposed to $T$), it was shown that if the superpotential dependence on the inflaton does not extend beyond $W_I$, from the canonical mass matrix, it is straightforward to see that there are no decay terms for the inflaton to scalars or fermions \cite{EKOTY,EGNO4}. 
This result can be readily seen if one performs the necessary field redefinitions to canonical fields. By defining K\"ahler normal coordinates (KNC), we can rewrite the Lagrangian and read off all possible couplings of the inflaton \cite{swamp1,swamp2}.  

We follow the same procedure as for the RNC. We consider a general theory with $N$ massive complex scalar fields denoted by $Z^I$ with general two-derivative interactions and a potential $V(Z, \bar{Z})$, given by the Lagrangian\footnote{We note that here we use barred and unbarred indices instead of the upper and lower indices as in Eq.~(\ref{Lkin3J}).}
\begin{equation}
    \label{lagcomplex}
    \mathcal{L} \; = \; K_{I \bar{J}}(Z, \bar{Z}) \partial_{\mu} Z^I \partial^{\mu} \bar{Z}^{\bar{J}} - V(Z, \bar{Z}) \, ,
\end{equation}
where $K_{I \bar{J}}$ is a metric tensor that is Hermitian. Here the use the holomorphic (unbarred) indices on the left and anti-holomorphic (barred) indices on the right, with $K^{I \bar{J}}K_{M \bar{J}} = \delta^I_M$. We expand the complex fields as
\begin{equation}
    \label{vaccomplex}
    Z^I \; = \; w^I + z^I, \qquad \bar{Z}^{\bar{I}} \; = \; \bar{w}^{\bar{I}} + \bar{z}^{\bar{I}} \, ,
\end{equation}
where $w^I$ is the complex scalar field VEV and $z^I$ is the field fluctuation. We introduce the complex tetrads that flatten the K\"ahler metric
\begin{equation}
    \label{comvielbconst}
    K_{I \bar{J}} (w, \bar{w}) e_\alpha{}^I(w,\bar{w}) e_{\bar{\beta}}{}^{\bar{J}}(w,\bar{w}) \; = \; \delta_{\alpha \bar{\beta}} \, .
\end{equation}
When we use the complex notation, the Greek indices are used for the mass eigenbasis, and the indices can be raised by using the Kronecker delta $\delta^{\alpha \bar{\beta}}$ and $\delta_{\alpha \bar{\beta}}$. The inverse complex tetrad $e^{\alpha}_{~I}(w, \bar{w})$ can be computed using the relation
$e^{\alpha}_{~I}(w, \bar{w}) = \delta^{\alpha \bar{\beta}} e_{\bar{\beta}}^{~\bar{J}}K_{I \bar{J}}$.

We introduce the covariant derivatives acting on the scalar potential $V(Z, \bar{Z})$:
\begin{equation}
    \label{covariantscomplex}
    V_{I_1 \ldots I_k}(w, \bar{w}) \; = \; \nabla_{I_1 \ldots} \nabla_{I_k} V(Z, \bar{Z})|_{w, \bar{w}} \, , \quad \partial_I V(Z, \bar{Z})|_{w, \bar{w}}  \; = \; 0 \, ,
\end{equation}
where these covariant derivatives are not symmetrized and we take the derivatives with respect to the complex fields $Z^i$ and $\bar{Z}^{\bar{I}}$.
Following Ref.~\cite{Wess:1992cp}, the non-zero Christoffel symbols are given by
\begin{equation}
    \label{eq:complchrist}
    \Gamma^{I}_{J K} \; = \; g^{I \bar{L}} g_{K \bar{L}, J}, \qquad \Gamma^{\bar{I}}_{\bar{J} \bar{K}} \; = \; g^{L \bar{I}} g_{L \bar{K}, \bar{J}} \, ,
\end{equation}
and the scalar mass matrix can be expressed as
\begin{equation}
    M^2 \; = \; 
    \begin{pmatrix}
        V_{I \bar{J}} & V_{I J} \\
        V_{\bar{I} \bar{J}} & V_{\bar{I} J}
    \end{pmatrix} \, .
\end{equation}
From the K\"ahler potential~(\ref{v1}), we find the following K\"ahler metric
\begin{equation}
    \begin{aligned}
    &K_{I \bar{J}} \; = \; \frac{1}{\left(T + T^* - \frac{|\phi|^2}{3} - \frac{|H_u|^2}{3} - \frac{|H_d|^2}{3} \right)^2} \times \\
   & \begin{pmatrix}
        3& -\phi& -H_u& -H_d \\
        -\phi^*& T+ T^* - \frac{|H_u|^2}{3} - \frac{|H_d|^2}{3}&\frac{H_u \phi^*}{3}& \frac{H_d \phi^*}{3}  \\
        -H_u^*& \frac{\phi H_u^*}{3}& T +T^* - \frac{|\phi|^2}{3} - \frac{|H_d|^2}{3}& \frac{H_d H_u^*}{3} \\
        -H_d^*& \frac{\phi H_d^*}{3} & \frac{H_u H_d^*}{3}& T+ T^* - \frac{|\phi|^2}{3} - \frac{|H_u|^2}{3}
    \end{pmatrix}\, .
    \end{aligned} 
\end{equation}
At the minimum, the VEVs of the complex fields $Z = (T, \phi, H_u, H_d)$ are given by $w = (\frac{1}{2},0,0,0)$, and the complex tetrad can be expressed as
\begin{equation}
    e_{\alpha}^{I}(w, \bar{w}) \; = \; {\rm diag}\left(\frac{1}{\sqrt{3}},1,1,1 \right) \, .
\end{equation}
If we compute the effective scalar potential~(\ref{effpot}) with the Cecotti superpotential~(\ref{cec}) combined with $W_{SM} = \mu H_u H_d$, and use the covariant derivatives~(\ref{covariantscomplex}) in the mass eigenbasis, 
\begin{equation}
    V_{I \bar{J} K}(w, \bar{w}) e_i^I(w, \bar{w}) e_{\bar j}^{\bar J}(w, \bar{w})  e_k^K(w, \bar{w}) \; = \; V_{i \bar{j} k}(w, \bar{w}) \, ,
\end{equation}
we find that the inflaton fluctuation $\delta T$ contains the following trilinear couplings to MSSM Higgs fields:\footnote{The complete set of couplings can be found in \cite{EGNO4}.}
\begin{equation}
    \mathcal{L} \; \supset \; \frac{\mu^2}{\sqrt{3}} \delta T |\delta H_u|^2 + \frac{\mu^2}{\sqrt{3}} \delta T |\delta H_d|^2 
    + \rm{h.c.} \, ,
\end{equation}
    where we have ignored a coupling of $T$ to $\phi^2$ as it does not lead to reheating. 
If instead we use the Wess-Zumino superpotential~(\ref{WWZ}), the inflaton fluctuation $\delta \phi$ does not have any trilinear couplings. Note these couplings are proportional to $\mu$. The decay rate is given by \cite{EGNO4,dgmo,dgkmo,kmo,kmov}
\beq
\Gamma = \frac{\mu^4}{12 \pi M M_P^2} \, ,
\eeq
for decays into the eight real scalar Higgs fields. For $\mu \sim \mathcal{O}(1)$~TeV, the reheating temperature is $\trh < .01$~eV. In the case of high scale supersymmetry (with $\mu \gtrsim m_\phi$), the reheating temperature can be significantly higher.

If we repeat the same procedure for the K\"ahler potential with twisted Higgs fields~(\ref{v2}), we find that the K\"ahler metric is given by
\begin{equation}
    K_{I \bar{J}} \; = \; 
    \begin{pmatrix}
        \frac{3}{\left(T+T^* - \frac{|\phi|^2}{3} \right)^2} &  -\frac{\phi}{\left(T+T^* - \frac{|\phi|^2}{3} \right)^2}&0&0 \\
         -\frac{\phi^*}{\left(T+T^* - \frac{|\phi|^2}{3} \right)^2} &  \frac{T+T^*}{\left(T+T^* - \frac{|\phi|^2}{3} \right)^2}&0&0 \\
         0&0&1&0\\
         0&0&0&1
    \end{pmatrix} \, .
\end{equation}
Using the same VEVs at the minimum as before, we find the complex tetrad
\begin{equation}
    e_{\alpha}^{I}(w, \bar{w}) \; = \; {\rm diag}\left(\frac{1}{\sqrt{3}},1,1,1 \right) \, .
\end{equation}
In this case for the Cecotti superpotential combined with $W_{SM} = \mu H_u H_d$, we find the following trilinear couplings to the MSSM Higgs for the inflaton fluctuation $\delta T$
\begin{equation}
    \mathcal{L} \; \supset \; \sqrt{3} \mu^2 \delta T |\delta H_u|^2 + \sqrt{3} \mu^2 \delta T |\delta H_d|^2 
    + \rm{h.c.} \, ,
\end{equation}
and for the Wess-Zumino model, there are no trilinear couplings for the inflaton fluctuation $\delta \phi$. These couplings, though 3 times larger, still provide a dismal amount of reheating. 

Note that there are of course many other potential couplings of the inflaton to MSSM scalars. For example, there is a three-body decay to a Higgs, stop, and antistop, with a decay rate $\mu^2 y_t^2 M/ 2048 \pi^3 M_P^2$ giving a reheating temperature of order 5 MeV \cite{EGNO4,building}. Four-body decay rates into stops (and anti-stops) are more promising and provide a rate $9 y_t^4 M^3/262144 \pi^5 M_P^2$, giving a reheating temperature of order $10^7$~GeV, provided the MSSM fields are in the twisted sector. For untwisted matter fields, this four-body rate vanishes. Including MSSM fermions, three-body decays into Higgs, top and antitop lead to the rate $3y_t^2M^3/2048\pi^3M_P^2$ for twisted fields, corresponding to $T_{\rm RH}\sim 10^9$~GeV. For more information, see, \cite{EGNO4,building}.

It is also possible that the inflaton does couple directly to SM fields if, for example, there is a superpotential coupling such as~\cite{ENO8,snu}
\beq
W_{\phi{\rm SM}} = y_\nu \phi H_u L \, ,
\eeq
to the up-like Higgs and lepton doublets. When considered
with the superpotential (\ref{WWZ}) there is an interaction term
$-M y_{\nu}H^*_u {\tilde{L}}^*\phi_1$, and
the inflaton decay width is given by
\beq\label{phisneu}
\Gamma(\phi \rightarrow H_u^0\tilde{\nu},H_u^+\tilde{f}_L) = M \frac{|y_{\nu}|^2}{16\pi}\, ,
\eeq
where we have neglected the masses of the final-state particles.
There is also a decay to fermions $\phi_1 \rightarrow H_u^0\tilde{\nu},H_u^+\tilde{f}_L$ with a rate equal to that in (\ref{phisneu}). Using Eq.~(\ref{deftrh}) for $\trh$ and
 the MSSM value $g_{\rm RH} = 915/4$ with $M =  3 \times 10^{13}$~GeV,
\beq
\trh = 4.8 \times 10^{14}~y_\nu~{\rm GeV} \, .
\eeq
Clearly, this is a very efficient way to reheat the Universe, if such a coupling exists.

It is also possible for the inflaton to decay to gauge bosons (and gauginos) if the gauge kinetic function $f_{\alpha\beta}$ has a non-trivial dependence on the inflaton \cite{EKOTY,klor,EGNO4,gkkmov}. In this case, the decay rate is given by\footnote{The numerical prefactor of this rate differs from the prefactor of the rate computed in~\cite{gkkmov} by a factor of 2. The reason is the definition of $d_g$ in terms of a derivative with respect to the complex field $\phi$ and not the physical inflaton, which is the canonically normalized real part, $\sqrt{2}\,{\rm Re}\,\phi$.}
\beq
\Gamma(\phi\rightarrow gg) = \Gamma(\phi \rightarrow \tilde{g}\tilde{g}) = \frac{d_{g}^2}{128\pi} N_G \frac{M^3}{M_P^2}\,,
\eeq
where $N_G = 12$ in the Standard Model (assuming a universal coupling of the inflaton to all gauge bosons), and $d_{g}$ is given by
\beq
d_{g} \equiv \langle {\rm Re}\,f\rangle^{-1}\left|\left\langle\frac{\partial f}{\partial \phi}\right\rangle\right| \, .
\eeq
This leads to a reheating temperature of 
\beq
T_{\rm RH} = 6.7 \times 10^{9}~d_{g }~{\rm GeV} \, .
\eeq

Finally, we note that the $T$-model potential for inflation found in Eq.~(\ref{EHTE2}), can also be derived in no-scale supergravity.
For Wess-Zumino-like models where the inflaton is $\phi$, a choice of superpotential \cite{GKMO1}
\beq
W = 2^{\frac{k}{4}+1}\sqrt{\lambda}  \left( \frac{\phi^{\frac{k}{2}+1}}{k+2} - \frac{\phi^{\frac{k}{2}+3}}{3(k+6)} \right)\, .
\eeq
Alternatively, choosing
\beq
W = \sqrt{\lambda}~\phi~(2T) \left(\sqrt{6} ~ \frac{2T-1}{2T+1} \right)^{\frac{k}{2}}\, ,
\eeq
yields the same potential when $T$ is associated with the inflaton.  

\section{Relating No-scale supergravity and $R + R^2$}
\label{sec:comp}

It is not coincidental that the Starobinsky potential can be derived from an $R + R^2$ theory of gravity as well as no-scale supergravity \cite{DLT,eno9}. The standard formulation of $N=1$ supergravity is in the Jordan frame, with $\mathcal{L} \supset -\frac16 \Phi R$ and only after a conformal transformation with $e^{2 \Omega} = \frac13 \Phi$ do we arrive at the theory defined in the Einstein frame with a K\"ahler potential given by 
\beq 
K = -6 \Omega = -3 \ln(\Phi/3) \, .
\label{phiK}
\eeq
More specifically, starting with \cite{cremmer}
\beq
{\mathcal L} = {\mathcal L}_{\rm aux} + {\mathcal L}^\prime \, ,
\eeq
where
\begin{eqnarray}
 {\mathcal L}_{\rm aux}  & = &  \frac{9}{\Phi} \left| \frac12 g^* + \cdots \right|^2  \frac{3}{\Phi J_{\phi\phi^*} } \left| \frac12 g^* \left( \frac{g^*_{\phi^*}}{g^*} - J_{\phi^*} \right) + \cdots \right|^2  \nonumber \\
 & - & \frac{1}{4\Phi} \left[ ( \Phi_{\phi^*} \partial_\mu \phi^* - \Phi_\phi \partial_\mu \phi) + \cdots \right]^2 
 \label{Laux} \, , \\
 {\mathcal L}^\prime & = &  - \Phi_{\phi \phi^*} |\partial_\mu \phi|^2 - \frac16 \Phi R \ + \cdots \, ,
\label{Lprime}
\end{eqnarray}
and 
$g$ is a holomorphic function of the scalars $\phi$ (indices on the scalars $\phi^i$ have been dropped for clarity) and $J = 6 \Omega = 3\ln(\Phi/3)$ is a function of $\phi, \phi^*$. This expression only includes the purely scalar part of the Lagrangian and the gravitational curvature. 
Collecting the kinetic terms in Eqs.~(\ref{Laux}) and (\ref{Lprime}), we have
\beq
{\mathcal L}_{\rm kin} = - \frac12 R - \left( \frac{3}{\Phi} \Phi_{\phi \phi^*} - \frac{3}{\Phi^2} \Phi_\phi \Phi_{\phi^*} \right) |\partial_\mu \phi|^2 =  - \frac12 R - J^j_i \partial_\mu \phi^i \partial^\mu \phi^*_j \, .
\label{Lkin2}
 \eeq
Setting $K = -J$, we see that this corresponds to the first two terms in Eq.~(\ref{Lkin3J}). 
The remaining terms in $\mathcal{L}_{\rm aux}$
can be reorganized to give the scalar potential given in Eq.~(\ref{effpot}), with $g = 2 W$. The simplest no-scale theory defined in Eq.~(\ref{CFKN}) is obtained when $\Phi = 3 (T + T^*)$ or  $3 (T + T^* - \phi \phi^*)$ in the more general theory with additional chiral fields as in Eq.~(\ref{v0}).

Notice now that the Starobinsky model can be matched to the real part of the supergravity theory with the identification of  $(1+ 2\alpha \Phi)$ from Eq.~(\ref{conftransf}) to $\Phi/3$ in Eq.~(\ref{phiK}). The latter can then be identified with $T+T^*$ to obtain the K\"ahler potential in Eq.~(\ref{CFKN}). 

If we include matter fields in the Starobinsky model as in Eq.~(\ref{model2}), where $H$ is a representative example of the matter fields $\phi^i$, we obtain the K\"ahler potential in Eq.~(\ref{v0}) if $\xi = 1/6$.  In fact, this was to be expected as we have seen that the supergravity couplings of the inflaton lead to vanishing decay rates from kinetic terms 
which is precisely the case in the $R + R^2$ model when matter fields are conformally coupled with $\xi = 1/6$ as in Eq.~(\ref{ratea2ein}).

We have seen that the conformal transformation to the Einstein frame in the  $R + R^2$ model
gives rise to a potential
\beq
V(\Phi) = \frac{\alpha}{2} \frac{\Phi^2} {(1 + 2 \alpha \Phi)^2} \, ,
\eeq
which becomes the Starobinsky potential 
when we set $\phi = \sqrt{3/2} \ln (1 + 2 \alpha \Phi)$. 
When matter fields are included, the potential is
\beq
V(\Phi) = \frac{\alpha}{2} \frac{\Phi^2} {(1 + 2 \alpha \Phi - 2 \xi |H|^2)^2} + \frac{V(H)}{(1 + 2 \alpha \Phi - 2 \xi |H|^2)^2} \, ,
\eeq
and now we must set $\phi = \sqrt{3/2} \ln (1 + 2 \alpha \Phi -2 \xi |H|^2)$. The second term here again closely resembles the scalar potential in no-scale supergravity given by Eq.~(\ref{VJ}) when we make the identification of $V(H)$ to ${\hat V}$, $1 + 2 \alpha \Phi$ with $(T+T^*)$, $\xi = 1/6$, and $H$ to a more general $\phi^i$. In the supergravity case of course ${\hat V}$ must be determined by the superpotential as in Eq.~(\ref{effVJ}). To completely match the scalar potentials in the $R^2$ and no-scale theories, additional potential interactions need to be added to the action in Eq.~(\ref{model2}) once a superpotential has been specified. 

We have seen that matter fields introduced in the no-scale framework appear as conformally coupled fields with $\xi = 1/6$ accounting for the lack to decay channels beyond those proportional to powers of the scalar masses which breaks the conformal symmetry.  
Previously, we have also seen that inflaton decays are expected to occur though the trace anomaly, and we expect the same to be true in the context of supergravity.  Indeed, the explicit factor of $e^{\frac23 K}$ in Eq.~(\ref{VJ})
is nothing other than the conformal factor $e^{-4\Omega} = e^{-2\sqrt{\frac23} \phi}$ in Eq.~(\ref{eq:acteinst}), with $K=-6\Omega$. The classical couplings of $K$ to scalar potential and kinetic terms may also be found from the coupling
\begin{align}
	\mathcal{L} = -\frac{K}{6} {T^\mu}_\mu\,,
 \label{KT}
\end{align}
as in Eq.~(\ref{OT}). We expect therefore that 
at the quantum level, there is a coupling of $K$ to $T^\mu_\mu|_{\rm anom}$ in Eq.~(\ref{anomT}). Thus up to a numerical factor,
we further expect that this coupling will induce a decay of the inflaton \cite{Endo:2007sz} as in Eq.~(\ref{anomD}).\footnote{This is only true when the inflaton is associated with $T$, as $\langle K_T \rangle \ne 0$ when evaluated at the minimum, in contrast to the case where the inflaton is $\phi$ and $\langle K_\phi \rangle = 0$.}
This coupling was also considered in the context of anomaly mediated supersymmetry breaking \cite{anom}.

\section{Summary}
\label{sec:sum}

Our view of inflation has evolved significantly since the original first order GUT phase transition proposed by Guth \cite{Guth:1980zm}.  The exit from accelerated expansion occurs naturally in slow-roll inflation. 
The problem of reheating is intimately connected with the detailed model of inflation and how the inflaton couples to the Standard Model. In addition, far from being simply an abstract construct employed to solve cosmological problems, inflation has become a theory with testable experimental predictions. Even simple models of inflation typically make three predictions:
The overall curvature, $\Omega = 1$; the tilt of the scalar anisotropy spectrum, $n_s \lesssim 1$, and the scalar-to-tensor ratio, r. From experiment, we have 
$\Omega = 0.9993 \pm 0.0019$ \cite{Planck}; $n_s  = 0.9649 \pm 0.0042$ \cite{Planck}; and $r < 0.036 $ 95\% CL \cite{rlimit} or $r < 0.032$ 95\% CL \cite{Tristram:2021tvh}. These values (and limits) can be compared for example to the predictions of the Starobinsky model \cite{Staro} which gives $\Omega = 1$, 
$n_s = 0.965$, and $r = 0.0035$. It should be noted that $n_s$ (and to a lesser extent $r$) depends on the number of e-folds which in turn depends on the reheating temperature \cite{egnov}. 

For this paradigm to work, reheating and the production of Standard Model particles must occur. Adding couplings
of the inflaton to the SM can in some cases become problematic if they distort the inflaton potential
and spoil the positive aspects of the inflationary expansion. In this work, we examined the question of inflaton decay and reheating in two related frameworks for inflation. When formulated as a modified theory of gravity with a Lagrangian given by $\frac12(-R + \alpha R^2)$ as in the Starobinsky model \cite{Staro}, we have seen that subsequent to the conformal transformation to the Einstein frame, couplings to Standard Model fields are automatically generated leading to a decay width proportional to $m_\phi^3/M_P^2$ and a reheating temperature of order $10^9$~GeV. This is the case so long as the Standard Model fields are not conformally coupled to curvature with conformal coupling $\xi = 1/6$ in which case the coupling of the inflaton to the Higgs field vanishes. 
However, even in this case, at the quantum level, there is a coupling of the inflaton to gauge fields through the trace anomaly leading to decay and a reheating temperature of order $10^8$~GeV.

We have also considered the inflationary models constructed in the framework of no-scale supergravity \cite{no-scale}. Once a relatively simple superpotential is specified \cite{eno6,Cecotti, FeKR,EGNO2,others,EGNO3} (as in Eq.~(\ref{WWZ}) or Eq.~(\ref{cec})), a Starobinsky-like potential is generated
yielding the same predictions for the inflationary observables. In this case, unless the inflaton is directly coupled to the matter (e.g., by associating the inflaton with the right-handed sneutrino \cite{ENO8,snu}), inflaton decay is highly suppressed. 
This can be understood when relating the no-scale supergravity models to the $R + R^2$ models as was done in the previous section, and one can see that the two theories are related when Standard Model fields are in fact conformally coupled with $\xi = 1/6$ in the $R+R^2$ theory. We expect that in this case too, inflaton decay through the trace anomaly is possible. Alternatively, coupling the inflaton to gauge fields through the gauge kinetic function may also lead to inflaton decay and reheating. 

We expect the next significant test of these models will be available in the next round of CMB experiments which can probe the tensor to scalar ratio down to $0.001$
and should either confirm or exclude the type of models discussed here which predict $r \sim 0.003$.

\authorcontributions{All work presented was a collaborative effort of Yohei Ema, Marcos A. G. Garcia, Wenqi Ke, Keith A. Olive, and Sarunas Verner. 
}

\funding{The~work of Y.E. and K.A.O. is supported in part by DOE grant DE-SC0011842 at the University of
Minnesota. The work of M.A.G.G. is supported by the DGAPA-PAPIIT grant IA103123 at UNAM, the CONAHCYT “Ciencia de
Frontera” grant CF-2023-I-17, and the PIIF-2023 grant from Instituto de F\'isica, UNAM.
}

\dataavailability{There is no new data to be made available. 
}

\acknowledgments{We would like to thank E. Dudas for helpful comments.  The~work of Y.E. and K.A.O. is supported in part by DOE grant DE-SC0011842 at the University of Minnesota. The work of M.A.G.G. is supported by the DGAPA-PAPIIT grant IA103123 at UNAM, the CONAHCYT “Ciencia de
Frontera” grant CF-2023-I-17, and the PIIF-2023 grant from Instituto de F\'isica, UNAM. The work of S.V is supported in part by DOE grant DE-SC0022148 at the University of Florida.}

\reftitle{References}


\end{document}